\documentclass[conference]{IEEEtran}
\IEEEoverridecommandlockouts
\usepackage{cite}
\usepackage{amsmath,amssymb,amsfonts}
\usepackage{algorithmic}
\usepackage{graphicx}
\usepackage{textcomp}
\usepackage{xcolor}

\usepackage{comment}
\usepackage{algorithm}
\usepackage{amsmath}
\newtheorem{lemma}{{Lemma}}

\usepackage{subfigure}

\def\BibTeX{{\rm B\kern-.05em{\sc i\kern-.025em b}\kern-.08em
    T\kern-.1667em\lower.7ex\hbox{E}\kern-.125emX}}
\begin{document}

\title{Joint Design of Coding and Modulation for Digital Over-the-Air Computation} 

\author{\IEEEauthorblockN{Xin Xie\IEEEauthorrefmark{1}, Cunqinq Hua\IEEEauthorrefmark{1}, Jianan Hong\IEEEauthorrefmark{1},Yuejun Wei\IEEEauthorrefmark{2}}
\IEEEauthorblockA{\IEEEauthorrefmark{1}School of Cyber Science and Engineering, Shanghai Jiao Tong University, Shanghai, China\\
\IEEEauthorrefmark{2}School of Computer and Information Engineering, Shanghai Polytechnic University, Shanghai, China}
\IEEEauthorblockA{xiexin\_312@sjtu.edu.cn, cqhua@sjtu.edu.cn, hongjn@sjtu.edu.cn, yjwei@sspu.edu.cn}
}

\maketitle

\begin{abstract}
Due to its high communication efficiency, over-the-air computation (AirComp) has been expected to carry out various computing tasks in the next-generation wireless networks. However, up to now, most applications of AirComp are explored in the analog domain, which limits the capability of AirComp in resisting the complex wireless environment, not to mention to integrate the AirComp technique to the existing universal communication standards, most of which are based on the digital system. In this paper, we propose a joint design of channel coding and digital modulation for digital AirComp transmission to attempt to reinforce the foundation for the application of AirComp in the digital system. Specifically, we first propose a non-binary LDPC-based channel coding scheme to enhance the error-correction capability of AirComp. Then, a digital modulation scheme is proposed to achieve the number summation from multiple transmitters via the lattice coding technique. We also provide simulation results to demonstrate the feasibility and the performance of the proposed design. 
\end{abstract}


\section{Introduction}
The next-generation (6G) wireless network will be a multi-layer cloud-fog-edge-terminal ubiquitous computing network to support various computing tasks, e.g., artificial intelligence (AI) algorithms \cite{Wang_COMST_23}. However, considering the limited spectrum bandwidth of wireless channels and the weak computing capability of wireless devices, it is imperative to excavate the computing potential of wireless networks themselves. As a promising technique, over-the-air computation (AirComp) can automatically accomplish the computing task in the physical layer while receiving from the transmitters by utilizing the superposition property of a wireless multiple-access channel (MAC), which greatly reduces the communication and computing cost of wireless networks.

Many research efforts have been devoted to study the application scenarios of AirComp since the cornerstone work of B. Nazer and M. Gastpar\cite{Nazer_TIT_07}. 
In \cite{Nazer_TIT_11}, AirComp was applied to the relay-assisted network by proposing a compute-and-forward relaying scheme, where the relay decodes a linear combination of all messages instead of ignoring interference as noise. In \cite{Xie_TMC_23}, AirComp was proposed for blockchain consensus in wireless networks to reduce communication overhead and computational complexity. AirComp also has been extended for model aggregation in the federated learning (FL) \cite{Yangkai_TWC_19, Zhu_TWC_21}, which is shown to reduce the communication latency without significant loss of the learning accuracy \cite{ZhuGX_WC_21}.

The channel superposition of AirComp can be achieved either in the analog domain or the digital domain. To resist non-ideal factors such as receiving noise and imperfect channel compensation, it is worth deploying the AirComp in the digital domain with error-correction capability. For instance, the authors in \cite{You_TMC_2023} have demonstrated that a coded digital AirComp system outperforms its analog counterpart in a  federated learning system. However, due to the lack of a matched digital modulation scheme for AirComp, the superposed symbols at the receiver are not codewords in \cite{You_TMC_2023}. Consequentially, the receiver has to decode the received symbols from the perspective of transmitter. That is, the decoder has to traverse all possible codeword combinations from all transmitters to search the most likely combination. Therefore, the decoding complexity exponentially increases with respect to the transmitter's number, which is tremendous when massive transmitters participate in the computing tasks.

To efficiently achieve a reliable digital AirComp system, we propose a joint design of channel coding and digital modulation scheme in this paper. Specifically, we first propose a non-binary LDPC-based channel coding scheme to enhance the error-correction capability of the AirComp system. Considering that the AirComp system requires all transmitters to align their channel fading, the summated signal from multiple transmitters can be equivalent to a single-source signal that is emitted by a virtual transmitter with summated source information. In this case, the AirComp system can be equivalent to an ordinary point-to-point system and a linear error-correcting code can be adopted. Then, we propose a digital modulation scheme for the AirComp system based on the lattice coding technique so that the superposed symbol at the receiver can be properly mapped to the natural summation of source information from all transmitters. Because the received symbols are also the codewords, the proposed decoding scheme only focuses on the received codewords instead of the combination of transmitted codewords, which significantly decreases the decoding complexity when compared with the decoding scheme in \cite{You_TMC_2023}.  

The rest of this paper is organized as follows. In Section~\ref{sec::system_model}, we introduce the system model of digital AirComp. We propose the channel coding and digital modulation schemes for the AirComp system in Section \ref{sec::coding} and Section~\ref{sec::modulation}, respectively. Simulation results are provided in Section \ref{sec::performance}. Finally, we summarize this work in Section \ref{sec::conclusion}.

\section{System Model}\label{sec::system_model}

We consider a wireless network with one base station (BS) and $K$ transmitters. In the system, all transmitters send signals to the BS through the same uplink channel. Ignoring the impact of receiving noise, the received signal at the BS is expressed as:
\begin{equation}\label{eq1}
	y=\sum_{k=0}^{K-1}h_k b_k x_k,
\end{equation}
where $h_k$, $b_k$, and $x_k$ is the channel fading, pre-processing factor, and transmitted signal from transmitter $k$ to the BS, respectively.  

Assuming that each transmitter $k$ has perfect knowledge of  $h_k$ and completely compensates it via the pre-processing as $b_k=1/h_k$. In this way, the signals transmitted by all transmitters will be aggregated at the BS as follows:
\begin{equation}\label{eq2}
	y= \sum_{k=0}^{K-1}x_k, 
\end{equation}
which computes the natural summation of all $x_k$s over the air.

Due to this property, the AirComp technique can be generalized to support a wide range of mathematical operations, which are based on the property of \emph{Nomographic} function in the following form:
\begin{equation}\label{eq3}
	f(s_0,s_1,...,s_{K-1})=\psi{\bigg(}\sum_{k=0}^{K-1}\phi{(s_k)}{\bigg)},
\end{equation}
where $\phi(\cdot)$ and $\psi(\cdot)$ denote pre- and post-processing functions, respectively. 

\begin{figure}[tb]
	\centering
	\includegraphics[scale=0.45]{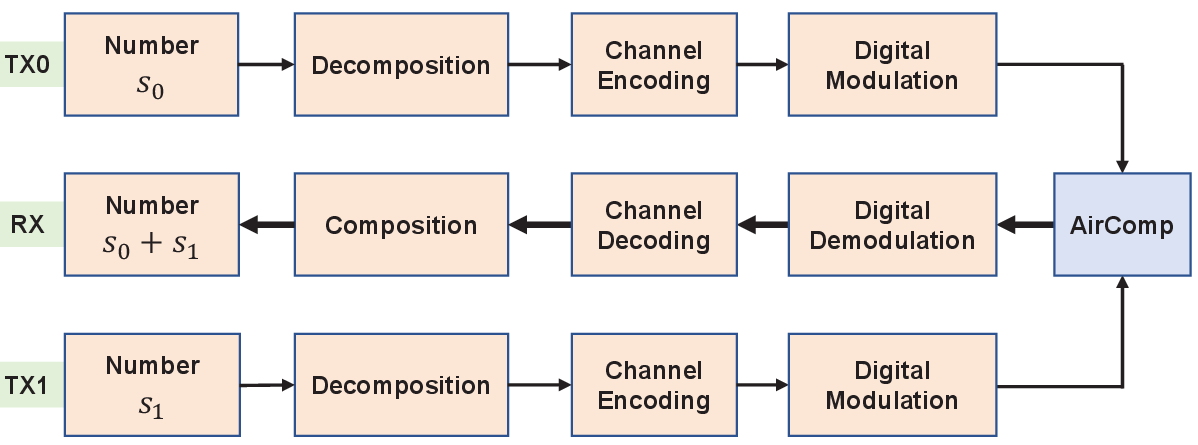}
	\caption{The overall procedures for digital AirComp system}
	\label{fig_system_model}
\end{figure}

The overall transmitting-receiving procedures for the AirComp system in the digital domain are present in Fig.~\ref{fig_system_model}. To achieve the natural summation of two numbers, three pairs of procedures will be carried out on the transmission link. First, considering that the number is usually large, e.g., using 16 bits or 32 bits to represent a number, it is necessary to decompose the number into multiple components that can be transmitted in the practical system. In the existing communication system, the number is usually represented by a binary vector. However, in the AirComp system, the summation operation may result in a carry, we consider a more general non-binary representation in this paper. 

In general, any number $s\in[0,\mathcal{S}-1]$ can be expressed by a small number $p$ as follows:
\begin{equation}\label{eq28}
    s = s_{l-1}p^{l-1}+s_{l-2}p^{l-2}+\cdots+s_{1}p+s_{0} = \mathbf{p}^T\cdot\mathbf{s}^{vec},
\end{equation}
where $\mathbf{s}^{vec}=[s_{l-1},s_{l-2},\cdots,s_0]^T, s_i\in\mathbb{Z}_p,\forall i\in[0,l-1]$, $\mathbf{p}=[p^{l-1},p^{l-2},\cdots,p,1]^T$ and $l=\lfloor{\log_p^\mathcal{S}}\rfloor+1$. Then we define the decomposition function $\mathcal{F}_\mathcal{S}(s,p,l): \mathbb{Z}_\mathcal{S}\mapsto\mathbb{Z}_p^l$ for $s$ by:
\begin{equation}\label{eq30}
    \mathcal{F}_\mathcal{S}(s,p,l)=\mathbf{s}^{vec}.
\end{equation}

Correspondingly, the composition function of $\mathbf{s}^{vec}$ can be given by:
\begin{equation}\label{eq31}
    \mathcal{G}_\mathcal{S}(\mathbf{s}^{vec},p,l)=\mathbf{p}^T\cdot\mathbf{s}^{vec}=s.
\end{equation}

For the AirComp system, the $\mathbf{s}^{vec}$ of each transmitter are transmitted and superposed at the receiver. The natural summation of $s_0$ and $s_1$ in Fig.~\ref{fig_system_model} can be easily proved by:
\begin{equation}\label{eq29}
    s_{\Sigma}=\mathcal{G}_\mathcal{S}(\mathbf{s}_0^{vec}+\mathbf{s}_1^{vec},p,l)=\mathbf{p}^T\cdot\mathbf{s}_0^{vec}+\mathbf{p}^T\cdot\mathbf{s}_1^{vec}=s_0+s_1.
\end{equation}

In the practical system, the decomposition result $\mathbf{s}^{vec}$ of source information $s$ will be sent to the channel coding module to increase the redundancy of the source and enhance the error correction capability. In this paper, we use the non-binary LDPC (NB-LDPC) code as the basic channel coding scheme. The LDPC codewords are then transformed into the field of real numbers in the digital modulation module so that they can be transmitted and superposed in the wireless channel. The details of channel coding and digital modulation will be described in Section \ref{sec::coding} and Section \ref{sec::modulation}, respectively. 

\section{Channel Coding Scheme}\label{sec::coding}
The main idea of channel coding for AirComp is to treat the multiple transmitters as one virtual transmitter based on the fact that the channel fading of all transmitters is aligned at the receiver. Then, the multi-transmitter AirComp system can be transformed into a point-to-point communication system and an existing channel coding scheme for a point-to-point system can be used as long as the scheme is linear. In the following, we study the scheme from both the encoding aspect and the decoding aspect.
\subsection{Encoding}
It is well known that the channel coding technique is performed on the finite field. 
For instance, the LDPC coding scheme can be performed in the finite field $\mathbb{Z}_q$ where $q$ is a prime. Specifically, each LDPC codeword element $v$ must belong to the interval $[0,q-1]$ and the result of multiplication/addition operation among elements needs to carry out the modulo-$q$ operation so that it also falls onto $[0,q-1]$. However, the target of AirComp is to achieve the natural addition, the modulo-$q$ operation will lead to a computing mistake. One feasible solution is to limit the natural addition result of AirComp data under the finite field order $q$, by which the natural addition and modulo-$p$ addition are equivalent. This solution can be mathematically described by the following lemma:
\begin{lemma}\label{lemma1}
    If $K$ transmitters participate in the computing task, and the information element of all transmitters falls onto the interval $[0,p-1]$, then the encoding scheme can be carried out in the finite field $\mathbb{Z}_q$ where $q$ is a prime and satisfies:
    \begin{equation}\label{eq11}
        K(p-1)\le q-1.
    \end{equation}
\end{lemma}	

After solving the modulo operation problem, we can obtain the encoding scheme of the AirComp system as follows: 

\subsubsection{Step 1} Generating the parity-check matrix $\mathbf{H}\in\mathbb{Z}^{M\times N}_q$, where $M$ is the number of parity-check constraints, $N$ is the length of codeword. The element of $\mathbf{H}$ in the $m$-th row and $n$-th column is denoted by $H_{m,n}$. Then the length of information elements in each codeword is $B = N-M$.

\subsubsection{Step 2} The $k$-th transmitter generates the information vector $\mathbf{s}_k=[s_{k,0},s_{k,1},\cdots,s_{k,B-1}]^T\in\mathbb{Z}_p^B$.

\subsubsection{Step 3} The information vector $\mathbf{s}_k$ is forwarded to the NB-LDPC encoder to calculate the codeword $\mathbf{c}_k\in\mathbb{Z}^{N}_q$ satisfying $\mathbf{H}\mathbf{c}_k=\mathbf{0}$. If we use a systematic encoding scheme, the codeword can be expressed by $\mathbf{c}_k=[s_{k,0},s_{k,1},\cdots,s_{k, B-1},g_{k,0},g_{k,1},\cdots,g_{k, M-1}]^T$=$[\mathbf{s}_k,\mathbf{g}_k]^T$, where $ \mathbf{g}_k=[g_{k,0},g_{k,1},\cdots,g_{k, M-1}]^T\in\mathbb{Z}_q^M$ is the parity-check vector.

\subsection{Decoding}

At the receiver, the modulo-$q$ summation of codewords can be given by:
\begin{equation}\label{eq12}
    \mathbf{c} = \underset{k=0}{\overset{K-1}{\bigoplus}}\mathbf{c}_k=\left[\sum_{k=0}^{K-1}\mathbf{s}_k,\underset{k=0}{\overset{K-1}{\bigoplus}}\mathbf{g}_k\right]^T,
\end{equation}
where $\bigoplus\mathbf{s}_k=\sum\mathbf{s}_k$ as long as the Lemma \ref{lemma1} is satisfied. We will present the modulation scheme that can achieve modulo-summation in the next section. Based on the fact that $\mathbf{H}\mathbf{c}_k=\mathbf{0}$, it is obvious that $\mathbf{H}\mathbf{c}=\bigoplus\mathbf{H}\mathbf{c}_k=\mathbf{0}$. Therefore, an ordinary NB-LDPC decoding algorithm can be performed to correct the potential transmission errors.

We take advantage of the common-used belief propagation (BP) algorithm in the log-domain to iteratively achieve the maximum a posteriori (MAP) performance. The first step is to calculate the log-likelihood ratio (LLR) of each element of $\mathbf{c}$ for the subsequent process. For the non-binary representation, we introduce the notion of LLR vector (LLRV) of variable $v$ over $\mathbb{Z}_q$ as follows:
\begin{equation}\label{eq18}
    \mathbf{L}(v)=[L(v=1),L(v=2),\cdots,L(v=q-1)]\in\mathbb{R}^{q-1},
\end{equation}
where
\begin{equation}\label{eq19}
    L(v=\alpha)=\ln{\frac{P(v=\alpha)}{P(v=0)}}, \alpha\in[1,q-1].
\end{equation}
Here $P(v=\alpha)$ denotes the probability that $v$ takes on the value $\alpha$. In the proposed scheme, the decoder should obtain the LLRV $\mathbf{L}_{ch}(c_n)$ of each element of codeword $\mathbf{c}=[c_0,c_1,\cdots,c_{N-1}]^T$ from the output of the demodulator. To successfully accomplish the decoding process, we further introduce the calculation of LLRV for the linear combination of two variables $v_1$ and $v_2$ with $A_1, A_2\in[1,q-1]$ as follows:
\begin{equation}\label{eq20}
    \mathbf{L}(A_1v_1+A_2v_2)=\boxplus (\mathbf{L}_1,\mathbf{L}_2,A_1,A_2),
\end{equation}
where $\boxplus (\mathbf{L}_1,\mathbf{L}_2, A_1, A_2)$ is the box-plus operator defined in \cite{Wymeersch_ICC_2004}, which can be recursively computed by comparison, addition and table look-up. Due to the space limitation, the details of box-plus is omitted in this paper. Then the detailed decoding scheme is present as follows \cite{Wymeersch_ICC_2004}:
\subsubsection{Initialization} For $m,n$ satisfying $H_{m,n}\ne 0$, initializing LLRV as follows:
\begin{equation}\label{eq13}
    \begin{aligned}
        \mathbf{L}(m\leftarrow n) &= \mathbf{L}_{ch}(c_n),\\
        \mathbf{L}(m\rightarrow n) &= \mathbf{0}.
    \end{aligned}
\end{equation}
Here $\mathbf{L}(m\leftarrow n)$ denotes the message transmitted from variable node $n$ to check node $m$, $\mathbf{L}(m\rightarrow n)$ denotes the message transmitted from check node $m$ to variable node $n$.

\subsubsection{Update variable node} The set of variable nodes connected to check node $m$ can be denote by an ordered set $\mathcal{N}(m)$ whose elements satisfy $n_{m,0}<n_{m,1}<\cdots<n_{m,|\mathcal{N}(m)|-1}$, where $|\mathcal{N}(m)|$ denotes the size of $\mathcal{N}(m)$. Further, we define new variables $\sigma_{m,n_{m,l}}=\sum_{j\le l}H_{m,n_{m,j}}c_{n_{m,j}}$ and $\rho_{m,n_{m,l}}=\sum_{j\ge l}H_{m,n_{m,j}}c_{n_{m,j}}$. Then it is obvious to prove that the value of $\sigma_{m,n_{m,l}}$ and $\rho_{m,n_{m,l}}$ can be computed recursively as follows:
\begin{equation}\label{eq14}
    \begin{aligned}
        \mathbf{L}(\sigma_{m,n_{m,l}}) &= \mathbf{L}(\sigma_{m,n_{m,l-1}}+H_{m,n_{m,l}}c_{n_{m,l}}),\\
        \mathbf{L}(\rho_{m,n_{m,l}}) &= \mathbf{L}(\rho_{m,n_{m,l+1}}+H_{m,n_{m,l}}c_{n_{m,l}}),
    \end{aligned}
\end{equation}
where the LLRV of $c_{n_{m,l}}$ is given by $\mathbf{L}(m\leftarrow n_{m,l})$. Then the message from check node $m$ to variable node $n_{m,l}$ is given by:
\begin{equation}\label{eq15}
    \begin{aligned}
        &\mathbf{L}(m\rightarrow n_{m,k}) =\\ &\mathbf{L}(-H^{-1}_{m,n_{m,k}}\sigma_{m,n_{m,k-1}}-H^{-1}_{m,n_{m,k}}\rho_{m,n_{m,k+1}}).
    \end{aligned}
\end{equation}

\subsubsection{Update check node} We denote $\mathcal{M}(n)$ as the set of check nodes connected to variable node $m$. Then the message from variable node $n$ to check node $m\in\mathcal{M}(n)$ is given by: 
\begin{equation}\label{eq16}
    \mathbf{L}(m\leftarrow n) = \mathbf{L}_{ch}(c_n)+\sum_{j\in \mathcal{M}(n)\backslash m}\mathbf{L}(j\rightarrow n).
\end{equation}

\subsubsection{Tentative decoding} Computing the a posteriori LLRV of each variable node:
\begin{equation}\label{eq17}
    \mathbf{L}_{post}(c_n) = \mathbf{L}_{ch}(c_n)+\sum_{j\in \mathcal{M}(n)}\mathbf{L}(j\rightarrow n).
\end{equation}

Based on $\mathbf{L}_{post}(c_n)$, the decoder can find the most likely value of $c_n$ that denoted by $\bar{c}_n$. Then constructing the vector $\bar{\mathbf{c}} = [\bar{c}_0,\bar{c}_1,\cdots,\bar{c}_{N-1}]^T$ and check the parity constraints by $\mathbf{z}=\mathbf{H}\bar{\mathbf{c}}$. If $\mathbf{z}=\mathbf{0}$, the decoding process is stopped; Otherwise, continue to carry out decoding process until reaching the maximum iteration number.

\subsection{Complexity analysis}
\begin{figure}[t]
	\centering
	\includegraphics[scale=0.17]{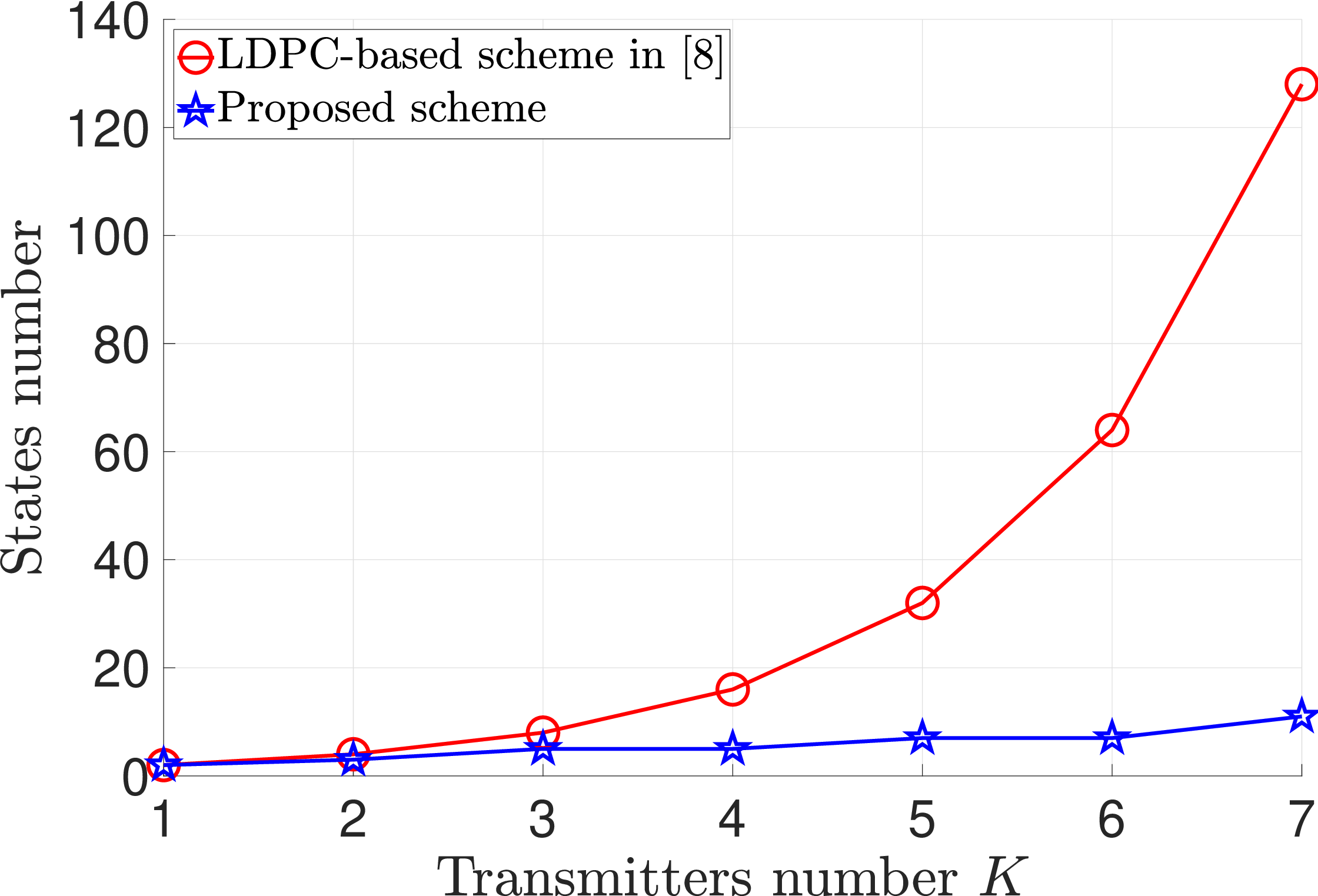}
	\caption{The decoding complexity, $p=2$}
	\label{fig_complexity}
\end{figure}

The essence of decoding is to compute the probability of all states that the codeword can arrive at and find the most likely one based on the received signal. In the proposed decoding scheme, each element of the received codewords has $q$ possible states. Therefore, according to Lemma \ref{lemma1}, the decoding complexity is at the order of $\mathbb{O}(pK)$. On the other hand, the decoding scheme in \cite{You_TMC_2023} finds the most likely state from the perspective of transmitters. Consequentially, the decoding complexity is at the order of $\mathbb{O}(p^K)$. Therefore, the proposed scheme has a significant advantage in the decoding complexity when the transmitter's number is large. In Fig.~\ref{fig_complexity}, we plot the state number of both decoding schemes with $p=2$ for $K$ varying from 1 to 7. The tendency of the curve also demonstrates our analysis.

\section{Digital Modulation Scheme}\label{sec::modulation}
In this section, we present the digital modulation scheme so that the LDPC codewords can be transmitted and summated in the practical system. Before that, we first give some fundamental knowledge about the lattice coding techniques

\subsection{Lattice coding}
A \emph{lattice} is an infinite discrete set of points in the Euclidean space that are regularly arranged and are closed under addition. A $D$-dimensional lattice $\boldsymbol{\Lambda}$ in the Euclidean space $\mathbb{R}^D$ can be generated as follows:
\begin{equation}\label{eq4}
	\boldsymbol{\boldsymbol{\Lambda}}=\{\mathbf{Gu}\ :\ \textbf{u}\in\mathbb{Z}^D\},
\end{equation}
where $\mathbf{G}\in\mathbb{R}^{D\times D}$ is a full-rank generator matrix.

A lattice $\boldsymbol{\boldsymbol{\Lambda}}_C$ is \emph{nested} in some lattice $\boldsymbol{\boldsymbol{\Lambda}}_F$ if $\boldsymbol{\boldsymbol{\Lambda}}_C\subseteq  \boldsymbol{\boldsymbol{\Lambda}}_F$. In this case, $\boldsymbol{\Lambda}_F$ is the fine lattice, which defines the codes, while $\boldsymbol{\Lambda}_C$ is the coarse lattice, which is also known as the shaping lattice. Specifically, the \emph{nested lattice codebook} $\mathcal{L}$ is the set of all points of a fine lattice $\boldsymbol{\Lambda}_F$ that is within the fundamental Voronoi region $\mathcal{V}_C$ of a coarse lattice $\boldsymbol{\Lambda}_C$:
\begin{equation}\label{eq5}
	\mathcal{L}=\boldsymbol{\Lambda}_F\cap\mathcal{V}_C=\{\mathbf{x}\ :\ \mathbf{x}=\boldsymbol{\lambda}\bmod \boldsymbol{\Lambda}_C,\boldsymbol{\lambda}\in\boldsymbol{\Lambda}_F\},
\end{equation}
where the \emph{fundamental Voronoi region}, $\mathcal{V}_C$, of the lattice $\boldsymbol{\Lambda}_C$, is the set of all points in $\mathbb{R}^D$ that are closer to the zero vector than any other lattice point of $\boldsymbol{\Lambda}_C$:
\begin{equation}\label{eq6}
	\mathcal{V}_C=\{\mathbf{z}\ :\ ||\mathbf{z}||\le||\mathbf{z}-\boldsymbol{\lambda}||,\forall\boldsymbol{\lambda}\in\boldsymbol{\Lambda}_C,\mathbf{z}\in\mathbb{R}^D\}.
\end{equation}

A \emph{lattice quantizer} is a map that sends a point $\mathbf{y}$ to the nearest lattice point in Euclidean distance:
\begin{equation}\label{eq7}
	Q_{\boldsymbol{\Lambda}}(\mathbf{y})=\arg\underset{\boldsymbol{\lambda}\in{\boldsymbol{\Lambda}}}{\min}||\mathbf{y}-\boldsymbol{\lambda}||.
\end{equation}
Then the \emph{modulus} of $\boldsymbol{\Lambda}$ can be represented by the quantization error of $\mathbf{y}$ with respect to $Q_{\boldsymbol{\Lambda}}(\mathbf{y})$ as follows:
\begin{equation}\label{eq8}
	\mathbf{y}\bmod{\boldsymbol{\Lambda}} = \mathbf{y}-Q_{\boldsymbol{\Lambda}}(\mathbf{y}).
\end{equation}

For the $k$-th transmitter, a $D$-dimension nested lattice code $\mathbf{x}_k$ can be generated based on the aforementioned LDPC codeword $\mathbf{c}_k$ by a modulation function $\phi(\cdot)$. Then for a nested lattice codebook $\mathcal{L}$, the following property holds for all ${\bf{x}}_k\in\mathcal{L}$:
\begin{equation}\label{eq9}
	\sum_{k=0}^{K-1}{\mathbf{x}}_k\bmod{\boldsymbol{\Lambda}}_C\in\mathcal{L},
\end{equation}
that is,  the sum of lattice codes modulo the shaping lattice is a code itself. Due to this linearity preserving characteristic \cite{Nazer_TIT_11}, there exists a demodulation function $\psi(\cdot)$ that satisfies:

\begin{equation}\label{eq10}
	\psi{\bigg (}\sum_{k=0}^{K-1}\phi(\mathbf{c}_k)\bmod{\boldsymbol{\Lambda}_C}{\bigg )}=\underset{k=0}{\overset{K-1}{\bigoplus}}\mathbf{c}_k,
\end{equation}
which is in line with definition of Nomographic function in \eqref{eq3}, where $f(\mathbf{c}_0,\mathbf{c}_1,\cdots, \mathbf{c}_{K-1})=\bigoplus\mathbf{c}_k$. Therefore, the lattice coding technique can be adopted in an AirComp system for the computation of modulo-summation.

\subsection{Modulation}
The target of modulation is to design a proper digital mapping scheme so that the modulo-summation of LDPC codewords in \eqref{eq12} can be achieved at the receiver.  The linearity preserving characteristic of lattice coding technique in \eqref{eq10} implies that the nested lattice coding can achieve this target over the air via a proper design of the mapping function $\phi(\cdot)$ and the demapping function $\psi(\cdot)$. 

According to the generation method of a general lattice in \eqref{eq4}, the mapping function is mainly determined by the generation matrix $\mathbf{G}\in\mathbb{R}^{D\times D}$. In this paper, we use the identity matrix $\mathbf{I}_D$ as the generation matrix, which corresponds to a cubic lattice. In a practical system, it can take at most 2-dimensional information simultaneously, we then naturally set $D\le 2$. As a result, the high-dimension codeword needs to be transmitted in segments. Specifically, the codeword can be decomposed as multiple components $\mathbf{u}\in\mathbb{{Z}}_q^D$ as $\mathbf{c}_k=[\mathbf{u}_{k,0},\mathbf{u}_{k,1},\cdots,\mathbf{u}_{k ,I-1}]\in\mathbb{Z}^{N}_q$ and the number of components is $I=N/D$. In this case, the size of the lattice codebook is $q^D$. Moreover, we need to normalize the lattice so that the fundamental Voronoi region of the shappling lattice falls onto the interval $[-1/2,1/2)$ at each dimension, i.e., the coarse lattice 
are all integers. Finally, the mapping function of $\mathbf{u}_{k,i}\in\mathbb{Z}_q^{D}$ to the lattice code can be formulated as $\mathbf{x}_{k,i}=\phi(\mathbf{u}_{k,i})$, where $\phi: \mathbb{Z}_q^D\mapsto \mathbb{R}^D$ is as follows: 
\begin{equation}\label{eq21}
	\begin{aligned}
	\phi(\mathbf{u}_{k,i}) &=  (\mathbf{I}_D\mathbf{u}_{k,i}/q)_1,
	\end{aligned}
\end{equation}
where $(x)_1$ denotes the modulo-$1$ operation and the output of $(x)_1$ is in the interval $[-1/2,1/2)$.

We consider the OFDM-based communication system where $\mathbf{x}_{k, i}$ is transmitted at $i$-th sub-carrier of the $k$ transmitter. At the BS, the received signal at the $i$-th sub-carrier is given by: 
\begin{equation}\label{eq22}
    \mathbf{y}_i=\sum_{k=0}^{K-1}{\mathbf{x}_{k,i}}+\mathbf{w}_i,\,i\in[0,I-1],
\end{equation}
where $\mathbf{w}_i\sim \mathcal{N}(\mathbf{0},\sigma_w^2\mathbf{I}_D)$ is the receiving noise.

\subsection{Demodulation}
If we ignore the receiving noise, the received signal $\mathbf{y}_i$ is the natural summation of lattice codes from all transmitters, which implies some summated lattices are not the valid elements from codebook generated in \eqref{eq21}. We then carry out the lattice modulo operation on $\mathbf{y}_i$ as follows:
\begin{equation}\label{eq23}
    \mathbf{t}_i=\mathbf{y}_i\mod{\boldsymbol{\Lambda}_C}.
\end{equation}

According to the linearity preserving characteristic of lattice coding in \eqref{eq9} and \eqref{eq10}, $\mathbf{t}_i$ is always the valid element from the codebook and is also corresponding to the modulo-$q$ summation of $\mathbf{u}_{k, i}$, which is denoted by:
\begin{equation}\label{eq24}
    \mathbf{v}_i=\underset{k=0}{\overset{K-1}{\bigoplus}}\mathbf{u}_{k,i}.
\end{equation}

\begin{figure}[tb]
	\centering	\includegraphics[scale=0.2]{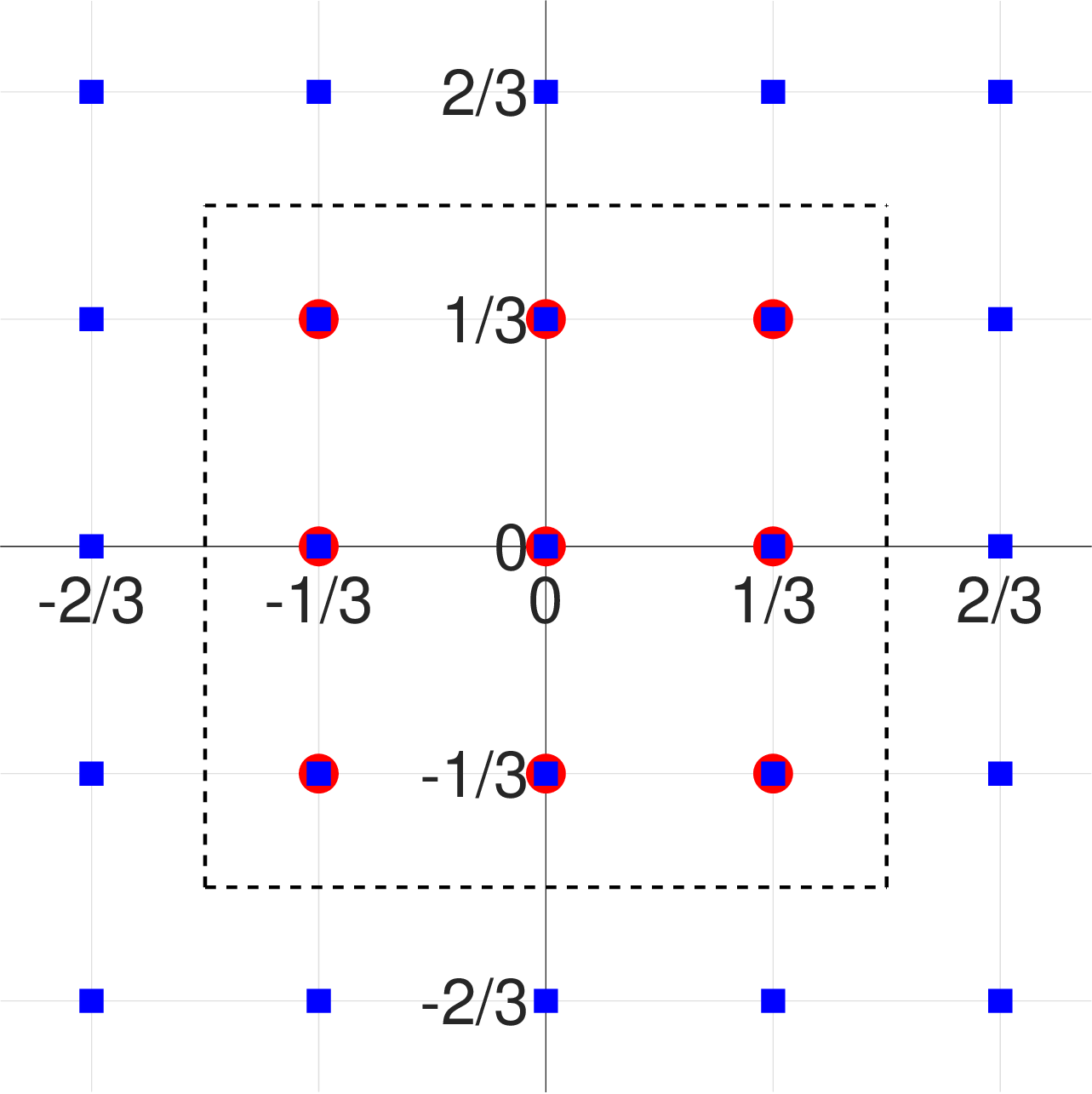}
	\caption{Nested $\mathbb{R}^2$ cubic lattice. Setting $p=2, D=2, K=2, q=3$. The 9 ($q^2$) red points are the valid lattice codes and the 25 blue squares are the possible received lattice points. The fundamental Voronoi region of the shaping lattice $\boldsymbol{\Lambda}_C$ is drawn in black dashed lines. With the lattice modulo operation, the received lattice points outside the fundamental Voronoi region of the shaping lattice fall inside the region and coincide with the valid lattice codes.}
	\label{fig_codeword}
\end{figure}

Therefore, the AirComp transmission scheme is transformed into a traditional point-to-point transmission scheme where the transmitted symbol is $\mathbf{v}_i$ and the received symbol is $\mathbf{t}_i$. For a better understanding, we present an example of the lattice code in Fig.~\ref{fig_codeword}.

Now we calculate the LLRV of $\mathbf{v}_i$ defined in \eqref{eq18} based on the received $\mathbf{t}_i$. It is known that $\mathbf{v}_i$ and $\mathbf{t}_i$ are scalar when $D=1$ and 2-dimensional real vector when $D=2$. For the general description, we denote $ \mathbf{v}_i=[v_i^{0},v_i^{D-1}]$ and $\mathbf{t}_i=[t_i^{0},t_i^{D-1}]$ in the rest of paper. Based on the MAP decoding criterion, the LLR of $v_i^{d}, d=\{0,1\}$ in \eqref{eq19} can be re-written as follows:
\begin{equation}\label{eq25}
    \begin{aligned}
        &L(v_i^{d}=\alpha) = \ln{\frac{P(v_i^{d}=\alpha|\mathbf{t}_i)}{{P(v_i^{d}=0|\mathbf{t}_i)}}}\\
        &=\ln{\frac{P(\mathbf{t}_i|v_i^{d}=\alpha)P(v_i^{d}=\alpha)}{{P(\mathbf{t}_i|v_i^{d}=0)P(v_i^{d}=0)}}}=\ln\frac{P(\mathbf{t}_i|v_i^{d}=\alpha)}{{P(\mathbf{t}_i|v_i^{d}=0)}}+\ln\Xi_{\alpha}.
        \end{aligned}
\end{equation}
where 
\begin{equation}\label{eq26}
    \begin{aligned}
        P(\mathbf{t}_i|v_i^{d}=\alpha)&=\sum_{\mathbf{x}_i: \mathbf{x}_i=\phi(
        \mathbf{v}_i), v_i^{d}=\alpha}P(\mathbf{t}_i|\mathbf{x}_i)\\
        &=\sum_{\mathbf{x}_i: \mathbf{x}_i=\phi(
        \mathbf{v}_i), v_i^{d}=\alpha}\frac{1}{\sqrt{2\pi}}exp(-\frac{1}{2\sigma_w^2}||\mathbf{t}_i-\mathbf{x}_i||^2).
    \end{aligned}
\end{equation}

Besides, the term $\Xi_\alpha = P(v_i^{d}=\alpha)/P(v_i^{d}=0)$ is the ratio of prior probability of $v_i^d$. We separate the calculation of $\Xi_\alpha$ into two parts: (1) For the $\mathbf{v}_i$ that corresponds to the summation of information vector $\mathbf{s}_k$, we generally assume that each element of $\mathbf{s}_k$ is equally distributed in the interval $[0,p-1]$, then the prior probability ratio of $\mathbf{v}_i$ can be calculated as follows:
\begin{equation}\label{eq27}
        \Xi_\alpha = \frac{P(v_i^{d}=\alpha)}{P(v_i^{d}=0)}=\frac{\sum_{\alpha_k: \sum \alpha_k=\alpha}\prod_k P(u_{k,i}^d=\alpha_k)}{\prod_k P(u_{k,i}^d=0)}=\Gamma(\alpha),
\end{equation}
where $\Gamma(\alpha)$ denotes the number of all possible combination of $\alpha_k\in[0,p-1]$ which satisfies $\sum_{k}\alpha_k=\alpha, \alpha\in[0,q-1]$. (2) The parity-check vector $\mathbf{g}_k$ is determined by the information vector $\mathbf{s}_k$ and the parity-check matrix $\mathbf{H}$. We assume the non-zeros elements of $\mathbf{H}$ are equally distributed in $[0,q-1]$, then each element of the $\mathbf{g}_k$ is equally distributed in the interval $[0,q-1]$. As a result, the $\mathbf{v}_i$ that corresponds to the summation of $\mathbf{g}_k$ is also equally distributed in the interval $[0,q-1]$ and the corresponding $\Xi_\alpha$ can be set as 1.

\section{Performance Evaluation}\label{sec::performance}
In this section, we evaluate the performance of the proposed scheme under different channel conditions and experimental setups.

\subsection{Simulation setup}
In the simulation, we adopt the $\mathbf{H}$ matrix in the WiFi standard (specifically, Table F-2 in \cite{wifi_80211_2012}) with the coding rate $1/2$. Considering that the proposed coding scheme is performed in the finite field $\mathbb{Z}_q$, we revise the original $\mathbf{H}$ matrix by replacing the non-zero elements of $\mathbf{H}$ with the random integer number that is equally distributed in the interval $[1,q-1]$. The codeword block length is 1296, therefore, the information block length is 648. By setting $l=6$ in \eqref{eq28}, a totally 108 number summation can be achieved in a codeword block. We adopt the block error rate (BLER) as the performance metric with $10^4$ random simulation runs. The decoding process is terminated after a maximum of 20 iterations.

\begin{figure*}
	\centering
	\subfigure[Different transmitting parameters, $K=2$.]{\includegraphics[scale=0.15]{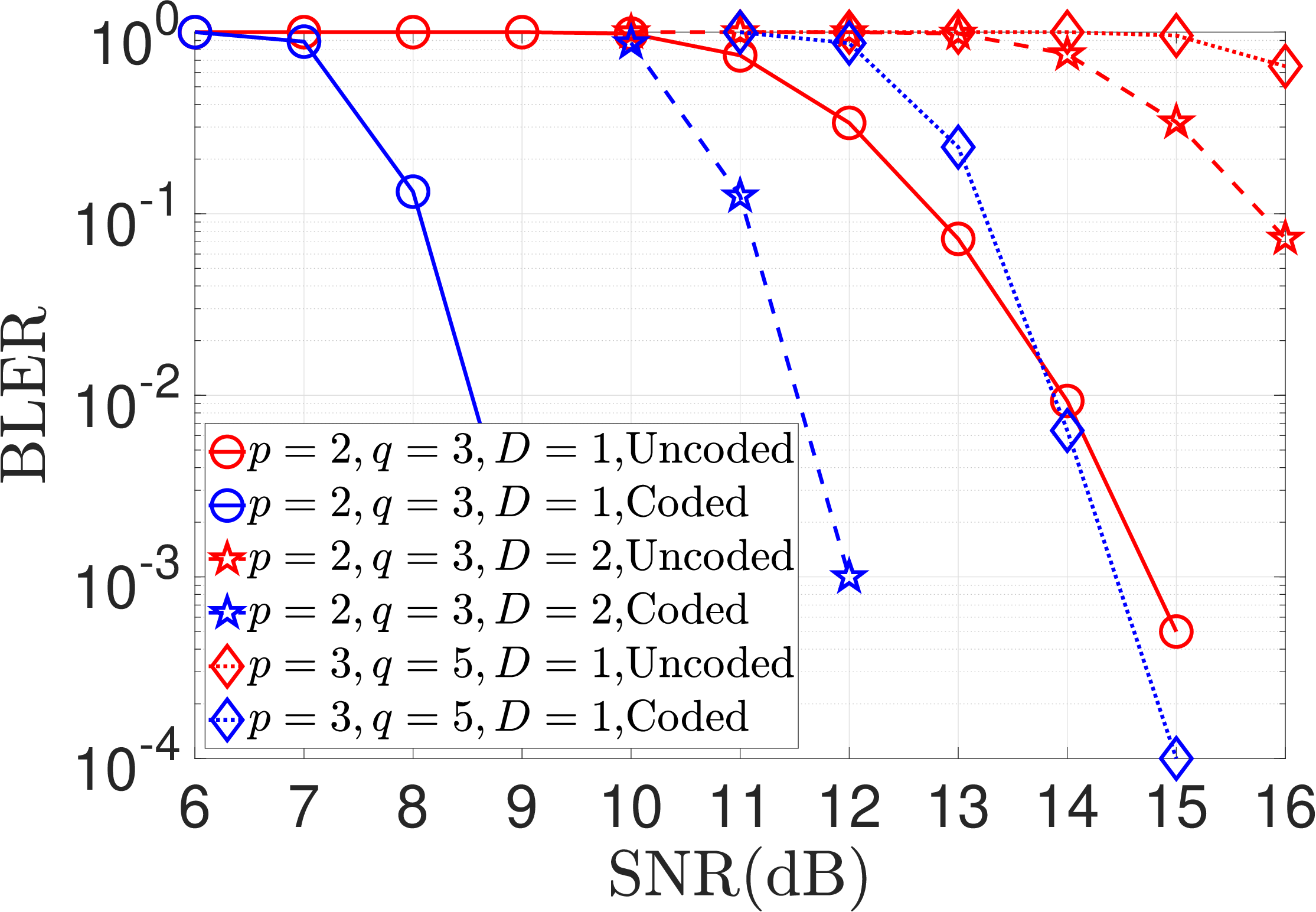}
	\label{fig_p_q}}
	\subfigure[Different transmitters number, $p=2, D=1$.]{\includegraphics[scale=0.15]{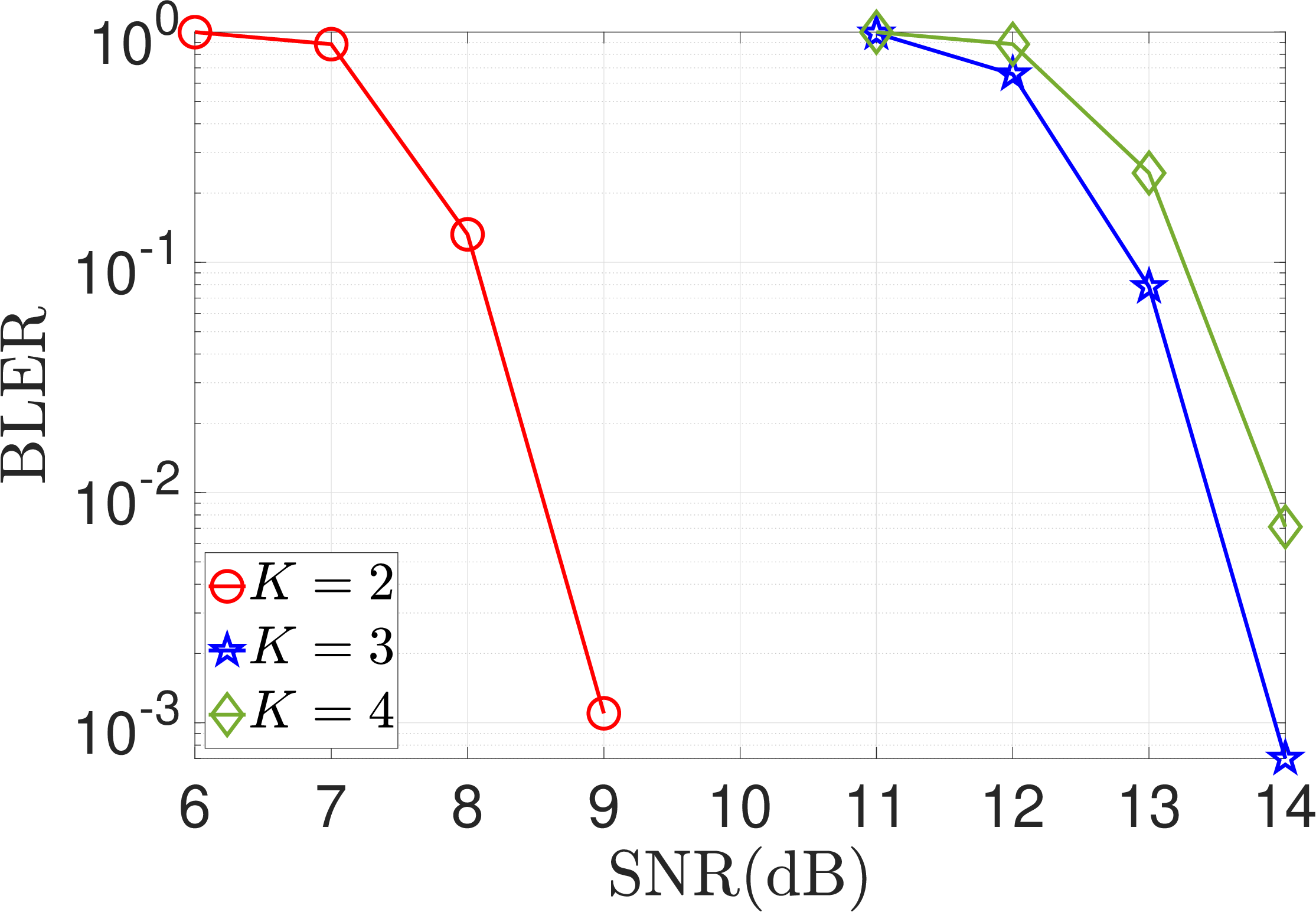}
	\label{fig_k}}
	\subfigure[Different channel phase shift.]{\includegraphics[scale=0.15]{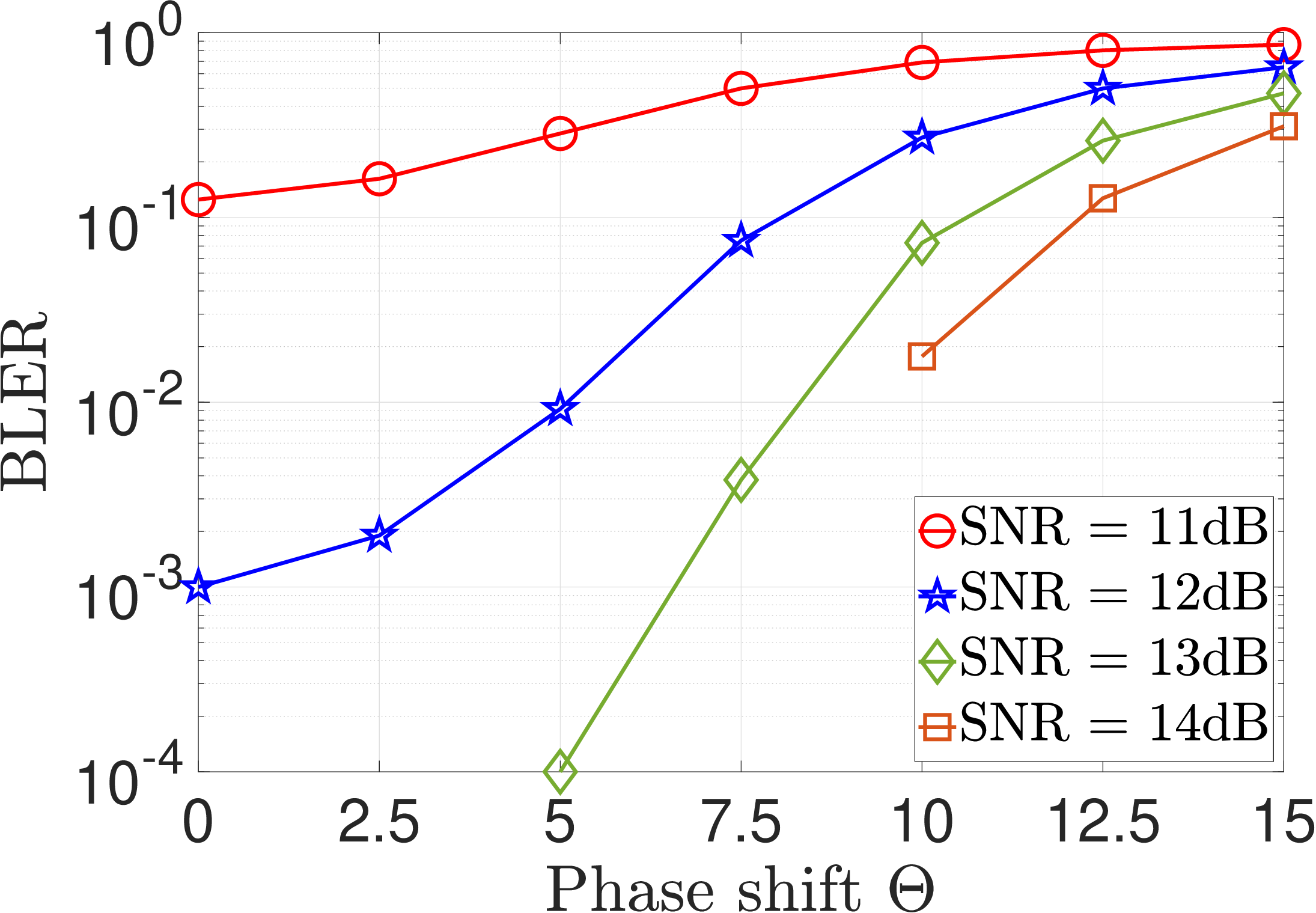}
	\label{fig_theta}}
	\caption{The performance of coded AirComp}
	\label{fig_Performance}
\end{figure*}

\subsection{Simulation results}
The error-correction capability of the proposed scheme is illustrated in Fig.~\ref{fig_p_q}. For the uncoded benchmarks, the proposed digital modulation scheme is also adopted to accomplish the summation task without channel coding usage. We can observe the significant performance improvements of coded AirComp schemes when compared with their uncoded counterparts. Moreover, it can be observed that the error-correction capability of the scheme with smaller $q$ is stronger. This is because the Euclidean distance of the codeword is larger. However, according to  Lemma \ref{lemma1}, the value of $q$ will limit the value of $p$ and $K$, which corresponds to the upper bound of the number that each transmitter can send (that is, $\mathcal{S}=p^l$) and the participating transmitters number $K$, respectively. Therefore, there exists a trade-off between the error-correction capability and computing efficiency and needs to be carefully considered in the practical system. 

We evaluate the performance of the proposed scheme under different number of transmitters in Fig.~\ref{fig_k}. We illustrate the performance curves when $K=\{2,3,4\}$ under a given $p=2$. According to the Lemma \ref{lemma1}, the value of $q$ is determined as follows: When $K=2$, $q=3$; When $K=3$ and $K=4$, $q=5$. Therefore, the performance of the former case is better than the latter two cases and the latter two cases have a similar performance. This result can be further verified by the case with $p=3, q=5, D=1$ in Fig.~\ref{fig_p_q}, which also has a similar performance curve as the cases when $K=3$ and $K=4$ in  Fig.~\ref{fig_k}. The above results demonstrate the system performance is mainly determined by the parameters at the receiver, e.g., $q$, rather than the parameters at the transmitter, e.g., $p$ or $K$.

Although the AirComp system works based on the requirement that all transmitters align their channel fading at the receiver, the actual channel compensation at the transmitter is always imperfect. We focus on the channel phase shift resulting from the residual CFO (carrier frequency offset) between the receiver and each transmitter. For each transmitter, a random channel phase shift is sampled with equal probability in the interval $[-\Theta,\Theta]$ and multiplied on the transmitted lattice code. We evaluate the system performance under different $\Theta$ in Fig.~\ref{fig_theta}. The increase of $\Theta$ implies that the transmitted symbol gets further away from the original code and can be more likely decoded to a mistaken code. Therefore, the BLER increases with the increase of $\Theta$. Essentially, the phase shift is one kind of noise, and a channel with better SNR condition has a higher tolerance for the phase shift. Therefore, the case with SNR=14dB has the best capability to resist the channel phase shift in this simulation setup. 

\section{Conclusion}\label{sec::conclusion}
In this paper, we presented a joint design of channel coding and digital modulation for the AirComp system, so that the AirComp technique can be seamlessly integrated into the universal digital communication systems. Compared with the existing works, the proposed scheme has lower operational complexity. This benefit comes from the fact that all transmitters should align their channel fading at the receiver. Therefore, the AirComp system can be transformed into an ordinary point-to-point system, and a channel coding scheme for the point-to-point system, such as NB-LDPC, can be directly adopted. Moreover, by using the lattice coding technique, the proposed digital modulation scheme can accomplish the modulo-summation at the symbols level so that channel decoding can be properly carried out in the finite field at the receiver. The simulation results are also provided to show the feasibility and the performance of the proposed design.

\bibliographystyle{IEEEtran}
\bibliography{AirCompCodingModulation_ICC}

\vspace{12pt}

\end{document}